\documentclass[conference]{IEEEtran}\usepackage[stretch=50,shrink=50]{microtype}

\usepackage{epsfig,amsmath,amssymb,epsf,amsthm,scalefnt,multirow,bm,mathtools,stfloats,stmaryrd}
\usepackage{xcolor}
\usepackage{float}
\usepackage{cite}
\usepackage{psfrag}
\usepackage{algorithm}
\usepackage{algpseudocode}

\newtheorem{lemma}{Lemma}
\newtheorem*{lemma*}{Lemma}

\newtheorem{corollary}{Corollary}

\algrenewcommand\algorithmicrequire{\textbf{Input:}}
\algrenewcommand\algorithmicensure{\textbf{Output:}}
\algdef{SE}[DOWHILE]{Do}{doWhile}{\algorithmicdo}[1]{\algorithmicwhile\ #1}

\begin{document}

\title{Performance of Cell-Free MmWave Massive MIMO Systems with Fronthaul Compression and DAC Quantization}

\author{\IEEEauthorblockN{In-soo Kim and Junil Choi}
\IEEEauthorblockA{School of Electrical Engineering\\
KAIST\\
Daejeon, Korea\\
E-mail: \{insookim, junil\}@kaist.ac.kr}}

\maketitle

\begin{abstract}
In this paper, the zero-forcing (ZF) precoder with max-min power allocation is proposed for cell-free millimeter wave (mmWave) massive multiple-input multiple-output (MIMO) systems using low-resolution digital-to-analog converters (DACs) with limited-capacity fronthaul links. The proposed power allocation aims to achieve max-min fairness on the achievable rate lower bounds of the users obtained by the additive quantization noise model (AQNM), which mimics the effect of low-resolution DACs. To solve the max-min power allocation problem, an alternating optimization (AO) method is proposed, which is guaranteed to converge because the global optima of the subproblems that constitute the original problem are attained at each AO iteration. The performance of cell-free and small-cell systems is explored in the simulation results, which suggest that not-too-small fronthaul capacity suffices for cell-free systems to outperform small-cell systems.
\end{abstract}

\section{Introduction}
Cell-free wireless communication systems make it possible to manage the resources across the cells at a geographical scale, thereby significantly enhancing the system performance \cite{7827017}. Moreover, millimeter wave (mmWave) massive multiple-input multiple-output (MIMO) boosts the system performance by offering large bandwidth and beamforming gain \cite{6894453}. The power consumption of cell-free mmWave massive MIMO systems, however, becomes impractical when a large number of radio frequency (RF) chains with high-resolution digital-to-analog converters (DACs) are deployed at the base stations. To reduce the power consumption, one possible solution is to employ hybrid architectures with low-resolution DACs \cite{8333733}.

Another potential bottleneck to the performance of cell-free systems is the constraint on the fronthaul capacity. In theory, the control unit and base stations in cell-free systems communicate through infinite-capacity fronthaul links. In practice, however, the fronthaul capacity is limited, so the advantage of cell-free systems cannot be fully exploited \cite{8678745, 8756286, 8891922}.

In this paper, the zero-forcing (ZF) precoder with max-min power allocation is proposed for cell-free mmWave massive MIMO systems with limited-capacity fronthaul links and low-resolution DACs. The goal of the proposed power allocation is to achieve max-min fairness on the achievable rate lower bounds of the users derived based on the additive quantization noise model (AQNM) \cite{9174405}, which mimics the effect of low-resolution DACs. We propose an alternating optimization (AO) method to solve the max-min power allocation problem. The convergence of the proposed AO method is shown to be guaranteed as the global optima of the subproblems that form the original problem are attained at each AO iteration. The performance of cell-free and small-cell systems is compared in the simulation results, which show that cell-free systems are energy-spectral efficient over small-cell systems provided that the fronthaul capacity is not too small.

\textbf{Notation:} $a$, $\mathbf{a}$, and $\mathbf{A}$ denote a scalar, vector, and matrix. $\mathbf{0}$ and $\mathbf{I}$ are an all-zero vector and identity matrix, whose dimensions are known from the context in a straightforward manner. The trace and determinant of $\mathbf{A}$ are $\mathrm{tr}(\mathbf{A})$ and $\mathrm{det}(\mathbf{A})$. The block diagonal matrix that contains $\mathbf{A}_{1}, \dots, \mathbf{A}_{n}$ as the block diagonal elements is $\mathrm{diag}(\mathbf{A}_{1}, \dots, \mathbf{A}_{n})$. The Kronecker product of $\mathbf{A}$ and $\mathbf{B}$ is $\mathbf{A}\otimes\mathbf{B}$. $\mathbf{A}\succ\mathbf{B}$ implies that $\mathbf{A}-\mathbf{B}$ is positive definite. $\mathbb{S}^{n\times n}$ is the set of $n\times n$ symmetric matrices. $I(\mathbf{x}; \mathbf{y})$ is the mutual information of random vectors $\mathbf{x}$ and $\mathbf{y}$.

\section{System Model}
Consider the downlink of a cell-free massive MIMO system with $M$ base stations and $K$ users where each base station has an $N_{\mathrm{BS}}$-antenna uniform planar array (UPA) and $N_{\mathrm{RF}}$ RF chains, while each user has an $N_{\mathrm{UE}}$-antenna UPA and one RF chain. The system operates in the mmWave band with the carrier frequency $f_{c}$ and bandwidth $W$. To reduce the power consumption, the base stations are equipped with a pair of $B$-bit DACs at each RF chain. The baseband processing is performed at a control unit, while the base stations act as radio units. In particular, the data signal is precoded at the control unit, compressed and conveyed through a fronthaul link limited to $C$ bps/Hz, and decompressed at the base station. In this section, the system model of the downlink of a cell-free mmWave massive MIMO system with limited-capacity fronthaul links and low-resolution DACs is developed.

\subsection{Fronthaul Compression and DAC Quantization}
In this subsection, the signal model that accounts for the fronthaul compression and DAC quantization is developed. The transmitted signal $\mathbf{x}_{m}\in\mathbb{C}^{N_{\mathrm{BS}}}$ of the $m$-th base station is
\begin{equation}\label{transmitted_signal}
\mathbf{x}_{m}=\mathbf{W}_{m}\mathrm{Q}\left(\vphantom{\sum_{k=1}^{K}\mathbf{f}_{m, k}\sqrt{\eta_{m, k}}s_{k}}\right.\underbrace{\sum_{k=1}^{K}\mathbf{f}_{m, k}\sqrt{\eta_{m, k}}s_{k}}_{=\mathbf{p}_{m}}+\mathbf{d}_{m}\left.\vphantom{\sum_{k=1}^{K}\mathbf{f}_{m, k}\sqrt{\eta_{m, k}}s_{k}}\right)
\end{equation}
where $s_{k}\sim\mathcal{CN}(0, 1)$ is the data signal transmitted to the $k$-th user, $\mathbf{f}_{m, k}\in\mathbb{C}^{N_{\mathrm{RF}}}$ and $\eta_{m, k}\geq 0$ are the linear precoder and power coefficient controlled by the control unit, $\mathbf{p}_{m}\in\mathbb{C}^{N_{\mathrm{RF}}}$ is the precoded signal to be transmitted at the $m$-th base station, and $\mathbf{W}_{m}\in\mathbb{C}^{N_{\mathrm{BS}}\times N_{\mathrm{RF}}}$ is the RF precoder of the $m$-th base station. The transmit power constraint is $P_{m}=\mathbb{E}\{\|\mathbf{x}_{m}\|^{2}\}\leq P$.

The fronthaul compression noise $\mathbf{d}_{m}\in\mathbb{C}^{N_{\mathrm{RF}}}$ represents the distortion when $\mathbf{p}_{m}$ is conveyed from the control unit to the base station through a fronthaul link limited to $C$ bps/Hz. In particular, $\mathbf{d}_{m}$ is modeled as $\mathcal{CN}(\mathbf{0}, \sigma_{m}^{2}\mathbf{I})$ where $\sigma_{m}$ suffices to satisfy \cite{8678745}
\begin{align}\label{fronthaul_capacity}
C_{m}&=I(\mathbf{p}_{m}; \mathbf{p}_{m}+\mathbf{d}_{m})\notag\\
     &=\log_{2}\det\left(\mathbf{I}+\frac{1}{\sigma_{m}^{2}}\sum_{k=1}^{K}\mathbf{f}_{m, k}\mathbf{f}_{m, k}^{\mathrm{H}}\eta_{m, k}\right)\leq C.
\end{align}
This assumption can be justified when the control unit and base station use large lattice codes to communicate through the fronthaul link \cite{508838, 6924850}. In essence, $C_{m}$ is the price to pay to convey $\mathbf{p}_{m}$ through the fronthaul link modeled as $\mathbf{p}_{m}+\mathbf{d}_{m}$.

The quantization function $\mathrm{Q}(\cdot)$ in \eqref{transmitted_signal} models $B$-bit DACs at the RF chains of the base stations, which quantizes the real and imaginary parts of each element with $2^{B}$ uniform quantization intervals that achieve the minimum mean squared error (MMSE) on Gaussian signals \cite{1057548}. To facilitate the analysis, the AQNM \cite{9174405} is adopted, which is a variant of the Bussgang theorem \cite{bussgang1952crosscorrelation} that linearizes the nonlinear distortion applied to Gaussian signals. Since $s_{k}$ and $\mathbf{d}_{m}$ are assumed to be Gaussian distributed, the AQNM can be applied to \eqref{transmitted_signal}, which yields
\begin{align}\label{linearized_transmitted_signal}
\mathbf{x}_{m}&=\mathbf{W}_{m}\left((1-\rho)\left(\sum_{k=1}^{K}\mathbf{f}_{m, k}\sqrt{\eta_{m, k}}s_{k}+\mathbf{d}_{m}\right)+\bar{\mathbf{q}}_{m}\right)\notag\\
              &=\mathbf{W}_{m}\left(\sum_{k=1}^{K}(1-\rho)\mathbf{f}_{m, k}\sqrt{\eta_{m, k}}s_{k}+\mathbf{q}_{m}\right)
\end{align}
where $\rho<1$ is the quantization distortion factor associated with the DAC resolution, $\bar{\mathbf{q}}_{m}\in\mathbb{C}^{N_{\mathrm{RF}}}$ is the quantization noise uncorrelated with $s_{k}$ and $\mathbf{d}_{m}$, and $\mathbf{q}_{m}\triangleq\bar{\mathbf{q}}_{m}+(1-\rho)\mathbf{d}_{m}\in\mathbb{C}^{N_{\mathrm{RF}}}$ is the aggregate noise uncorrelated with $s_{k}$. The exact values of $\rho$ are listed in \cite{1057548}, and $\rho$ can be approximated as $\rho\approx\pi\sqrt{3}/2\cdot 2^{-2B}$ when $B\gg 1$ \cite{9174405}. Since fronthaul links and low-resolution DACs operate independently across the base stations, $\mathbf{q}_{m}$ is independent across $m$. The second moment of $\bar{\mathbf{q}}_{m}$ is \cite{9174405}
\begin{equation}
\mathbf{C}_{\bar{\mathbf{q}}_{m}}=\rho(1-\rho)\mathrm{diag}\left(\sum_{k=1}^{K}\mathbf{f}_{m, k}\mathbf{f}_{m, k}^{\mathrm{H}}\eta_{m, k}+\sigma_{m}^{2}\mathbf{I}\right),
\end{equation}
so the second moment of $\mathbf{q}_{m}$ becomes
\begin{equation}\label{second_moment}
\mathbf{C}_{\mathbf{q}_{m}}=\rho(1-\rho)\mathrm{diag}\left(\sum_{k=1}^{K}\mathbf{f}_{m, k}\mathbf{f}_{m, k}^{\mathrm{H}}\eta_{m, k}\right)+(1-\rho)\sigma_{m}^{2}\mathbf{I}.
\end{equation}
The distribution of $\mathbf{q}_{m}$, however, is unknown because $\bar{\mathbf{q}}_{m}$ is non-Gaussian with some unknown distribution.

\subsection{Downlink Cell-Free MmWave Massive MIMO System}
In this subsection, the downlink model of a cell-free mmWave massive MIMO system is developed. The channel $\mathbf{H}_{k, m}\in\mathbb{C}^{N_{\mathrm{UE}}\times N_{\mathrm{BS}}}$ between the $m$-th base station and $k$-th user is formed by $L_{k, m}$ paths as
\begin{equation}
\mathbf{H}_{k, m}=\sum_{\ell=1}^{L_{k, m}}g_{k, m, \ell}\mathbf{a}_{\mathrm{UE}}(\theta_{k, m, \ell}^{\mathrm{UE}}, \phi_{k, m, \ell}^{\mathrm{UE}})\mathbf{a}_{\mathrm{BS}}(\theta_{k, m, \ell}^{\mathrm{BS}}, \phi_{k, m, \ell}^{\mathrm{BS}})^{\mathrm{H}}
\end{equation}
where $\mathbf{a}_{\mathrm{BS}}(\theta, \phi)\in\mathbb{C}^{N_{\mathrm{BS}}}$ and $\mathbf{a}_{\mathrm{UE}}(\theta, \phi)\in\mathbb{C}^{N_{\mathrm{UE}}}$ are the array response vectors of the base station and user. The $\ell$-th path is associated with the path gain $g_{k, m, \ell}\in\mathbb{C}$, horizontal angle-of-arrival (AoA) $\theta_{k, m, \ell}^{\mathrm{UE}}\in[-\pi, \pi]$, vertical AoA $\phi_{k, m, \ell}^{\mathrm{UE}}\in[0, \pi]$, horizontal angle-of-departure (AoD) $\theta_{k, m, \ell}^{\mathrm{BS}}\in[-\pi, \pi]$, and vertical AoD $\phi_{k, m, \ell}^{\mathrm{BS}}\in[0, \pi]$.

The received signal $y_{k}\in\mathbb{C}$ of the $k$-th user is
\begin{align}\label{received_signal}
y_{k}&=\mathbf{w}_{k}^{\mathrm{H}}\left(\sum_{m=1}^{M}\mathbf{H}_{k, m}\mathbf{x}_{m}+\mathbf{n}_{k}\right)\notag\\
     &=\sum_{m=1}^{M}\mathbf{h}_{k, m}^{\mathrm{H}}\left(\sum_{i=1}^{K}(1-\rho)\mathbf{f}_{m, i}\sqrt{\eta_{m, i}}s_{i}+\mathbf{q}_{m}\right)+v_{k}
\end{align}
where $\mathbf{w}_{k}\in\mathbb{C}^{N_{\mathrm{UE}}}$ is the RF combiner of the $k$-th user, and $\mathbf{n}_{k}\sim\mathcal{CN}(\mathbf{0}, N_{0}W\mathbf{I})$ represents the additive white Gaussian noise (AWGN) with power spectral density $N_{0}$. The effective channel $\mathbf{h}_{k, m}\triangleq\mathbf{W}_{m}^{\mathrm{H}}\mathbf{H}_{k, m}^{\mathrm{H}}\mathbf{w}_{k}\in\mathbb{C}^{N_{\mathrm{RF}}}$ and AWGN $v_{k}\triangleq\mathbf{w}_{k}^{\mathrm{H}}\mathbf{n}_{k}\in\mathbb{C}$ with variance $\sigma^{2}=N_{0}W\|\mathbf{w}_{k}\|^{2}$ are introduced for notational simplicity. In this paper, $\mathbf{W}_{m}$ and $\mathbf{w}_{k}$ are assumed to be fixed, which are controlled by the base stations and users. So, the goal is to maximize the achievable rates of the users in some sense subject to the transmit power constraint and \eqref{fronthaul_capacity} by controlling $\{\mathbf{f}_{m, k}\}_{\forall m, k}$ and $\{\eta_{m, k}\}_{\forall m, k}$ at the control unit. The control unit is assumed to have full channel state information (CSI) of the cell-free system.

\section{ZF Precoder with Max-Min Power Allocation}
In this section, the ZF precoder with max-min power allocation is proposed that accounts for the fronthaul compression and DAC quantization. To facilitate the analysis in the sequel, define
\begin{align}
&\mathbf{F}_{m}=\begin{bmatrix}\mathbf{f}_{m, 1}&\cdots&\mathbf{f}_{m, K}\end{bmatrix},\ \mathbf{F}^{\mathrm{H}}=\begin{bmatrix}\mathbf{F}_{1}^{\mathrm{H}}&\cdots&\mathbf{F}_{M}^{\mathrm{H}}\end{bmatrix},\notag\\
&\mathbf{h}_{k}^{\mathrm{H}}=\begin{bmatrix}\mathbf{h}_{k, 1}^{\mathrm{H}}&\cdots&\mathbf{h}_{k, M}^{\mathrm{H}}\end{bmatrix},\ \mathbf{H}^{\mathrm{H}}=\begin{bmatrix}\mathbf{h}_{1}&\cdots&\mathbf{h}_{K}\end{bmatrix},\notag\\
&\bm{\eta}_{m}=\mathrm{diag}(\eta_{m, 1}, \dots, \eta_{m, K}),\ \bm{\sigma}=\mathrm{diag}(\sigma_{1}, \dots, \sigma_{M})\notag\\
&\mathbf{q}^{\mathrm{H}}=\begin{bmatrix}\mathbf{q}_{1}^{\mathrm{H}}&\cdots&\mathbf{q}_{M}^{\mathrm{H}}\end{bmatrix}\notag
\end{align}
for notational convenience. Then, $y_{k}$ in \eqref{received_signal} can be re-expressed using $\mathbf{h}_{k}$ and $\mathbf{q}$ as
\begin{align}\label{reexpressed_received_signal}
y_{k}=&\sum_{i=1}^{K}\sum_{m=1}^{M}(1-\rho)\mathbf{h}_{k, m}^{\mathrm{H}}\mathbf{f}_{m, i}\sqrt{\eta_{m, i}}s_{i}+\mathbf{h}_{k}^{\mathrm{H}}\mathbf{q}+v_{k}\notag\\
     =&\sum_{m=1}^{M}(1-\rho)\mathbf{h}_{k, m}^{\mathrm{H}}\mathbf{f}_{m, k}\sqrt{\eta_{m, k}}s_{k}\notag\\
      &+\sum_{i\neq k}\sum_{m=1}^{M}(1-\rho)\mathbf{h}_{k, m}^{\mathrm{H}}\mathbf{f}_{m, i}\sqrt{\eta_{m, i}}s_{i}+\mathbf{h}_{k}^{\mathrm{H}}\mathbf{q}+v_{k}.
\end{align}
To eliminate the inter-user interference that corresponds to the second term in the last equality of \eqref{reexpressed_received_signal}, the ZF precoder is adopted from \cite{7917284}, which proceeds as follows. The power coefficients are forced to be dependent only on $k$, so the subscript $m$ is dropped from $\{\eta_{m, k}\}_{\forall m, k}$ as $\bm{\eta}=\mathrm{diag}(\eta_{1}, \dots, \eta_{K})$. Then, the linear precoders are configured to satisfy\footnote{Although imposing the unit norm constraint on the linear precoders as $\|\mathbf{f}_{m, k}\|=1$ is more conventional, we let the power coefficients to satisfy the transmit power constraint instead of normalizing the linear precoders without loss of generality.}
\begin{equation}
\sum_{m=1}^{M}\mathbf{h}_{k, m}^{\mathrm{H}}\mathbf{f}_{m, i}=\begin{cases}1\ \text{if}\ i=k\\0\ \text{if}\ i\neq k\end{cases},
\end{equation}
which is accomplished by setting $\mathbf{F}$ as the pseudoinverse of $\mathbf{H}$. As a result, the inter-user interference is eliminated from \eqref{reexpressed_received_signal} as
\begin{equation}
y_{k}=(1-\rho)\sqrt{\eta_{k}}s_{k}+\mathbf{h}_{k}^{\mathrm{H}}\mathbf{q}+v_{k}.
\end{equation}

The achievable rates of the users, however, have no closed form expressions because $\mathbf{q}$ is non-Gaussian with some unknown distribution as pointed out in \eqref{second_moment}. The achievable rate lower bounds of the users, in contrast, admit closed form expressions, which can be obtained using the worst-case noise assumption \cite{959289, 7931630}. To proceed, consider the second moment of $\mathbf{q}$, which is given as
\begin{align}
\mathbf{C}_{\mathbf{q}}&=\mathrm{diag}(\mathbf{C}_{\mathbf{q}_{1}}, \dots, \mathbf{C}_{\mathbf{q}_{M}})\notag\\
                       &=\rho(1-\rho)\mathrm{diag}(\mathbf{F}\bm{\eta}\mathbf{F}^{\mathrm{H}})+(1-\rho)\bm{\sigma}^{2}\otimes\mathbf{I}
\end{align}
where the block off-diagonal elements are zero because $\mathbf{q}_{m}$ is independent across $m$. Since the worst-case noise among the distributions with the same second moment is Gaussian, the achievable rate lower bound of the $k$-th user is obtained as
\begin{equation}
\log_{2}\left(\vphantom{\frac{(1-\rho)^{2}\eta_{k}}{\mathbf{h}_{k}^{\mathrm{H}}\mathbf{C}_{\mathbf{q}}\mathbf{h}_{k}+\sigma^{2}}}\right. 1+\underbrace{\frac{(1-\rho)^{2}\eta_{k}}{\mathbf{h}_{k}^{\mathrm{H}}\mathbf{C}_{\mathbf{q}}\mathbf{h}_{k}+\sigma^{2}}}_{=\mathrm{SQNR}_{k}}\left.\vphantom{\frac{(1-\rho)^{2}\eta_{k}}{\mathbf{h}_{k}^{\mathrm{H}}\mathbf{C}_{\mathbf{q}}\mathbf{h}_{k}+\sigma^{2}}}\right)\leq I(s_{k}; y_{k})
\end{equation}
where $\mathrm{SQNR}_{k}$ is the signal-to-quantization-plus-noise ratio (SQNR) of the $k$-th user assuming $\mathbf{q}\sim\mathcal{CN}(\mathbf{0}, \mathbf{C}_{\mathbf{q}})$.

In this paper, the power allocation that achieves max-min fairness on the achievable rate lower bounds of the users under the transmit power constraint and \eqref{fronthaul_capacity} is of primary interest, namely
\begin{alignat}{3}\label{p1}
&\max_{\bm{\eta}\succeq\mathbf{0}, \bm{\sigma}\succeq\mathbf{0}}\quad&&\min_{k}\mathrm{SQNR}_{k}\notag\\
&\mathrm{subject}\ \mathrm{to}\quad                                  &&P_{m}\leq P,\quad&&m=1, \dots, M,\notag\\
&                                                                    &&C_{m}\leq C,\quad&&m=1, \dots, M.\tag{P1}
\end{alignat}

To solve \eqref{p1}, $\mathrm{SQNR}_{k}$, $P_{m}$, and $C_{m}$ are expanded in matrix forms as follows. $\mathrm{SQNR}_{k}$ admits the form
\begin{align}\label{sqnr}
&\mathrm{SQNR}_{k}(\bm{\eta}, \bm{\sigma})=\notag\\
&\frac{(1-\rho)^{2}\eta_{k}}{\rho(1-\rho)\mathbf{h}_{k}^{\mathrm{H}}\mathrm{diag}(\mathbf{F}\bm{\eta}\mathbf{F}^{\mathrm{H}})\mathbf{h}_{k}+(1-\rho)\mathbf{h}_{k}^{\mathrm{H}}(\bm{\sigma}^{2}\otimes\mathbf{I})\mathbf{h}_{k}+\sigma^{2}},
\end{align}
whose numerator and denominator are linear functions of $\bm{\eta}$. Likewise, $P_{m}$ is obtained from \eqref{linearized_transmitted_signal} and \eqref{second_moment} in conjunction with the fact that $\mathbf{q}_{m}$ is uncorrelated with $s_{k}$ as
\begin{align}\label{ex}
P_{m}(\bm{\eta}, \sigma_{m})=&(1-\rho)^{2}\mathrm{tr}(\mathbf{W}_{m}\mathbf{F}_{m}\bm{\eta}\mathbf{F}_{m}^{\mathrm{H}}\mathbf{W}_{m}^{\mathrm{H}})\notag\\
                             &+\rho(1-\rho)\mathrm{tr}(\mathbf{W}_{m}\mathrm{diag}(\mathbf{F}_{m}\bm{\eta}\mathbf{F}_{m}^{\mathrm{H}})\mathbf{W}_{m}^{\mathrm{H}})\notag\\
                             &+(1-\rho)\sigma_{m}^{2}\mathrm{tr}(\mathbf{W}_{m}\mathbf{W}_{m}^{\mathrm{H}}).
\end{align}
Re-expressing \eqref{fronthaul_capacity}, $C_{m}$ is written as
\begin{equation}\label{c}
C_{m}(\bm{\eta}, \sigma_{m})=\log_{2}\mathrm{det}\left(\mathbf{I}+\frac{1}{\sigma_{m}^{2}}\mathbf{F}_{m}\bm{\eta}\mathbf{F}_{m}^{\mathrm{H}}\right).
\end{equation}
By inspection, the dependence of $\mathrm{SQNR}_{k}$, $P_{m}$, and $C_{m}$ on $\bm{\eta}$ and $\bm{\sigma}$ is deduced from \eqref{sqnr}, \eqref{ex}, and \eqref{c} as follows.
\begin{itemize}
\item $\mathrm{SQNR}_{k}$ is a quasilinear function of $\bm{\eta}$.
\item $P_{m}$ is a linear function of $\bm{\eta}$.
\item $C_{m}$ is a concave function of of $\bm{\eta}$.
\item $\mathrm{SQNR}_{k}$ is a non-increasing function of $\bm{\sigma}\succeq\mathbf{0}$.
\item $P_{m}$ is a non-decreasing function of $\sigma_{m}\geq 0$.
\item $C_{m}$ is a non-increasing function of $\sigma_{m}\geq 0$.
\end{itemize}
A function $f: \mathbb{S}^{n\times n}\to\mathbb{R}$ is said to be non-decreasing when $\mathbf{X}\preceq\mathbf{Y}$ implies $f(\mathbf{X})\leq f(\mathbf{Y})$. In general, \eqref{p1} is nonconvex. Using the properties above, an AO method that solves \eqref{p1} is proposed, whose convergence is shown to be guaranteed.

The proposed AO method alternates between $\bm{\eta}$ and $\bm{\sigma}$. In particular, two subproblems of \eqref{p1} are alternatingly solved, one solved over $\bm{\eta}$ for fixed $\bm{\sigma}$ and the other solved over $\bm{\sigma}$ for fixed $\bm{\eta}$. The convergence of the proposed AO method is guaranteed provided that the global optima of the two subproblems are attained at each AO iteration.

To update $\bm{\eta}$, the bisection method \cite{boyd2004convex} is adopted as proposed in \cite{7917284, 8678745}, which is an iterative algorithm where one bisection iteration is cast as a feasibility problem
\begin{alignat}{3}\label{p2}
&\mathrm{find}\quad                &&\bm{\eta}\succeq\mathbf{0}\notag\\
&\mathrm{subject}\ \mathrm{to}\quad&&\mathrm{SQNR}_{k}\geq t,\quad&&k=1, \dots, K,\notag\\
&                                  &&P_{m}\leq P,\quad            &&m=1, \dots, M,\notag\\
&                                  &&C_{m}\leq C,\quad            &&m=1, \dots, M\tag{P2}
\end{alignat}
with $t$ varying from iteration to iteration. The bisection method attains the global optimum of the subproblem over $\bm{\eta}$ at each AO iteration provided that \eqref{p2} is solved at each bisection iteration. The constraints in \eqref{p2}, however, are nonconvex because $C_{m}$ is concave with respect to $\bm{\eta}$. Therefore, the classical convex optimization approach \cite{7917284, 8678745} cannot solve \eqref{p2}.

To solve \eqref{p2}, Lemma \ref{lemma} is adopted from \cite{7917284}, which was originally established for cell-free systems with infinite-capacity fronthaul links and infinite-resolution DACs.
\begin{lemma}\label{lemma}
Consider two diagonal matrices $\bm{\eta}$ and $\bm{\eta}'\succ\mathbf{0}$ that satisfy $\mathrm{SQNR}_{k}(\bm{\eta}, \bm{\sigma})=t$ and $\mathrm{SQNR}_{k}(\bm{\eta}', \bm{\sigma})\geq t$ for $k=1, \dots, K$. Then, $\bm{\eta}'\succeq\bm{\eta}\succeq\mathbf{0}$ holds. \cite{7917284}
\end{lemma}
Lemma \ref{lemma} is still valid because Lemma \ref{lemma} was derived based on the assumption that the linear coefficients of $\bm{\eta}$ in the numerator and denominator of $\mathrm{SQNR}_{k}$ are non-negative, which indeed are as evident from \eqref{sqnr}. Then, \eqref{p2} can be solved using the following corollary.
\begin{corollary}\label{corollary}
\eqref{p2} is feasible if and only if the solution $\bm{\eta}$ of $\mathrm{SQNR}_{k}(\bm{\eta}, \bm{\sigma})=t$ for $k=1, \dots, K$ satisfies the constraints in \eqref{p2}.
\begin{proof}
Consider only ``only if" because ``if" is trivial. Since \eqref{p2} is feasible, there exists $\bm{\eta}'\succ\mathbf{0}$ that satisfies the constraints in \eqref{p2}. Therefore, the solution $\bm{\eta}$ of $\mathrm{SQNR}_{k}(\bm{\eta}, \bm{\sigma})=t$ for $k=1, \dots, K$ obeys $\bm{\eta}'\succeq\bm{\eta}\succeq\mathbf{0}$ by Lemma \ref{lemma}, which implies that $\bm{\eta}$ satisfies the non-negative definite constraint and first constraint with equality. Since $P_{m}$ is a linear function of $\bm{\eta}$ with non-negative linear coefficients as evident from \eqref{ex}, $P_{m}(\bm{\eta}, \sigma_{m})\leq P_{m}(\bm{\eta}', \sigma_{m})\leq P$ holds for $m=1, \dots, M$ due to $\bm{\eta}'\succeq\bm{\eta}$, which implies that $\bm{\eta}$ satisfies the second constraint. To check the third constraint, consider the expansion
\begin{align}
\underbrace{\mathbf{I}+\frac{1}{\sigma_{m}^{2}}\mathbf{F}_{m}\bm{\eta}'\mathbf{F}_{m}^{\mathrm{H}}}_{=\mathbf{A}+\mathbf{B}}=\underbrace{\frac{1}{\sigma_{m}^{2}}\mathbf{F}_{m}(\bm{\eta}'-\bm{\eta})\mathbf{F}_{m}^{\mathrm{H}}}_{=\mathbf{A}}+\underbrace{\mathbf{I}+\frac{1}{\sigma_{m}^{2}}\mathbf{F}_{m}\bm{\eta}\mathbf{F}_{m}^{\mathrm{H}}}_{=\mathbf{B}}.
\end{align}
Since $\mathbf{A}\succeq\mathbf{0}$ and $\mathbf{B}\succeq\mathbf{0}$ due to $\bm{\eta}'-\bm{\eta}\succeq\mathbf{0}$ and $\bm{\eta}\succeq\mathbf{0}$, applying Minkowski's inequality \cite{104312} yields $\mathrm{det}(\mathbf{A}+\mathbf{B})\geq\mathrm{det}(\mathbf{A})+\mathrm{det}(\mathbf{B})\geq\mathrm{det}(\mathbf{B})$. Therefore, $C_{m}(\bm{\eta}, \sigma_{m})\leq C_{m}(\bm{\eta}', \sigma_{m})\leq C$ holds for $m=1, \dots, M$ due to the monotonicity of logarithm, which implies that $\bm{\eta}$ satisfies the third constraint.
\end{proof}
\end{corollary}
Since the numerator and denominator of $\mathrm{SQNR}_{k}$ are linear functions of $\bm{\eta}$, finding the solution $\bm{\eta}$ of $\mathrm{SQNR}_{k}(\bm{\eta}, \bm{\sigma})=t$ for $k=1, \dots, K$ is equivalent to solving a set of $K$ linear equations. After the solution is obtained, Corollary \ref{corollary} suggests \eqref{p2} is solved by checking the feasibility of the solution.

To update $\bm{\sigma}$, note that decreasing $\bm{\sigma}$ increases $\mathrm{SQNR}_{k}$ for $k=1, \dots, K$, hence the objective function of \eqref{p1}. Furthermore, the first constraint becomes more feasible as $\bm{\sigma}$ decreases because $P_{m}$ is a non-decreasing function of $\sigma_{m}\geq 0$. Since $C_{m}$ is a non-increasing function of $\sigma_{m}\geq 0$, the global optimum of the subproblem over $\bm{\sigma}$ at each AO iteration is the one that satisfies the third constraint with equality. Since $C_{m}$ is differentiable with respect to $\sigma_{m}$, the solution can be found numerically without much effort using derivative-based root-finding algorithms, Newton's method, for example.

\section{Simulation Results}
In this section, the performance of cell-free mmWave massive MIMO systems with limited-capacity fronthaul links and low-resolution DACs is evaluated based on the achievable rate lower bounds of the users and per-user energy efficiency lower bound. The data signal is precoded using the proposed ZF precoder with max-min power allocation. The performance of small-cell systems with low-resolution DACs is considered as a baseline. In this paper, small-cell refers to systems where the base stations operate independently with no control unit, which means that the data signal is precoded at the base stations rather than a centralized control unit. Therefore, small-cell systems have no fronthaul capacity constraint. The base stations in small-cell systems serve $K/M$ nearest users. In addition, the maximum ratio transmission (MRT), ZF, and regularized ZF (RZF) precoders with full power transmission are considered for small-cell systems. For both cell-free and small-cell systems, the RF combiners of the users are set as the dominant left singular vector of the channel to the nearest base station. Then, the RF precoders of the base stations are set as the top-$N_{\mathrm{RF}}$ right singular vectors of the effective RF channel $\begin{bmatrix}\mathbf{H}_{k_{1}, m}^{\mathrm{H}}\mathbf{w}_{k_{1}}&\cdots&\mathbf{H}_{k_{K/M}, m}^{\mathrm{H}}\mathbf{w}_{k_{K/M}}\end{bmatrix}^{\mathrm{H}}$ to the nearest $K/M$ users. To satisfy the constant modulus constraint on the RF precoders and combiners, the alternating projection method \cite{1377501} is used at the base stations and users, which is commonly adopted in RF precoder design \cite{8057288, 8333733}. At each alternating projection iteration, the semi-unitary matrix is projected to the constant modulus space, which is again projected to the semi-unitary space.

The system parameters are $M=63$, $K=630$, $N_{\mathrm{BS}}=64$, $N_{\mathrm{RF}}=16$, $N_{\mathrm{UE}}=2$, $f_{c}=30$ GHz, $W=80$ MHz, $P=33$ dBm, and $N_{0}=-174$ dBm/Hz. The locations of the micro layer base stations are set according to the Dense Urban-eMBB scenario in ITU-R M.2412-0 \cite{itu2017guidelines}. Likewise, the channels and locations of the users are randomly generated as specified in the Dense Urban-eMBB scenario. The macro layer inter-site distance (ISD) is 200 m.

\begin{figure}[t]
\centering
\includegraphics[width=1\columnwidth]{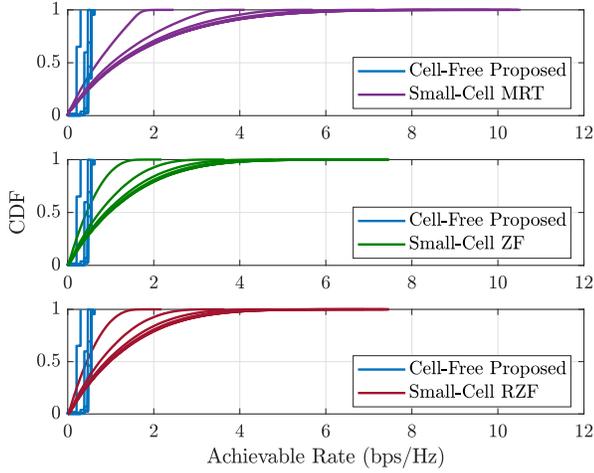}
\caption[caption]{The CDF of the achievable rate lower bounds of the users in cell-free and small-cell systems for $C=16$ bps/Hz. In each subplot, the lines with the same color correspond to $B=1, \dots, 8, \infty$ bits from left to right.}\label{figure_1}
\end{figure}

\begin{figure}[t]
\centering
\includegraphics[width=1\columnwidth]{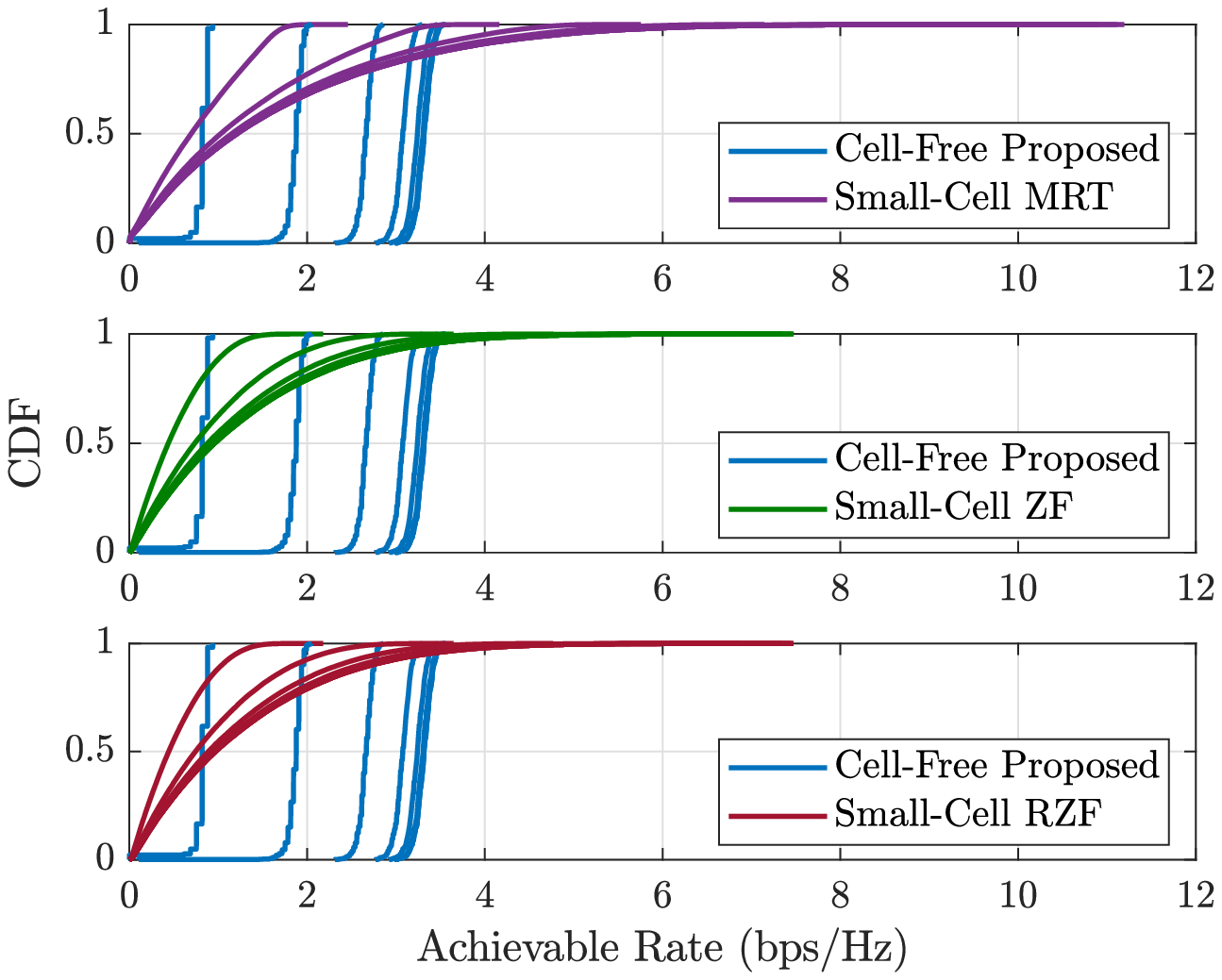}
\caption[caption]{The CDF of the achievable rate lower bounds of the users in cell-free and small-cell systems for $C=64$ bps/Hz. In each subplot, the lines with the same color correspond to $B=1, \dots, 8, \infty$ bits from left to right.}\label{figure_2}
\end{figure}

\begin{figure}[t]
\centering
\includegraphics[width=1\columnwidth]{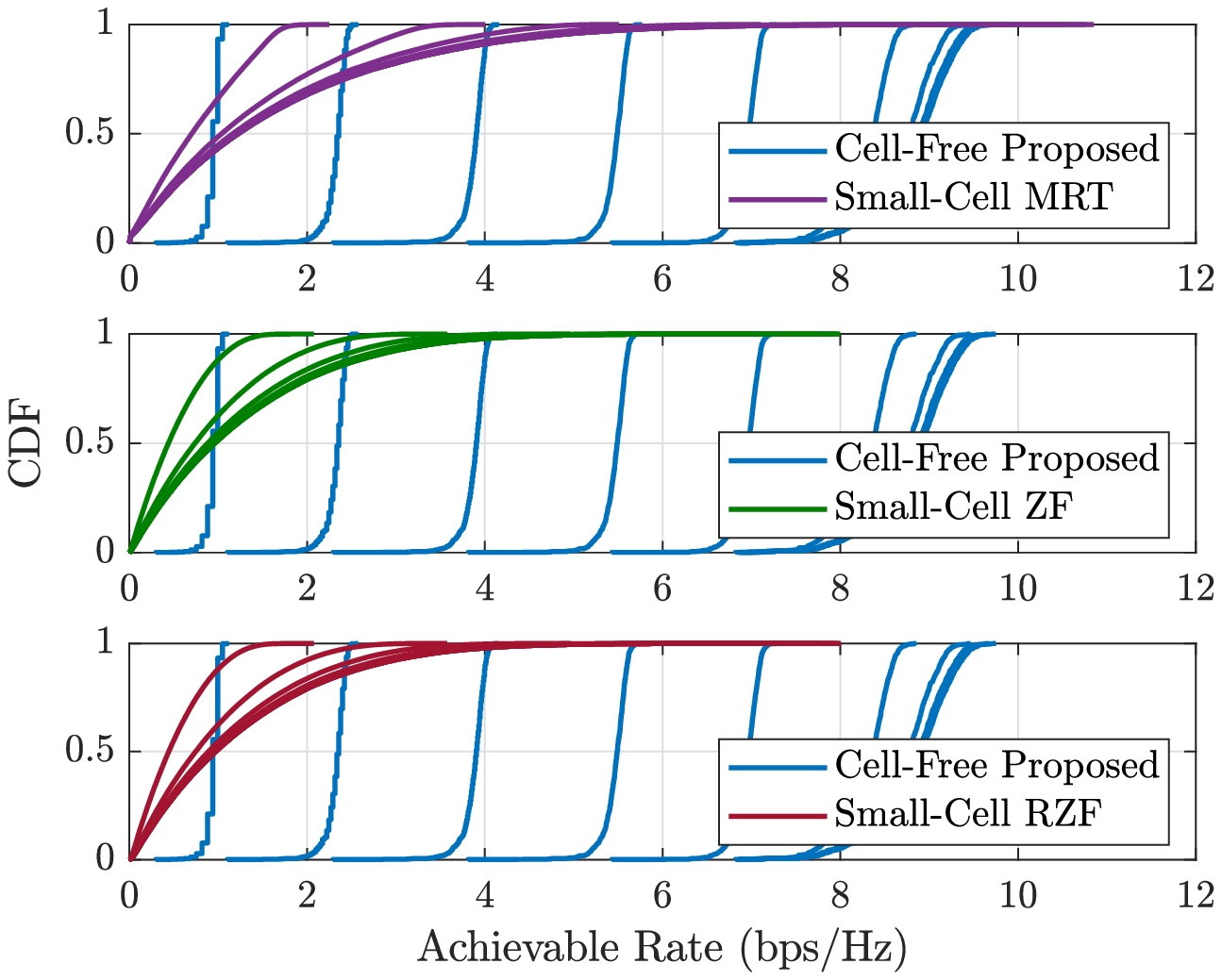}
\caption[caption]{The CDF of the achievable rate lower bounds of the users in cell-free and small-cell systems for $C=256$ bps/Hz. In each subplot, the lines with the same color correspond to $B=1, \dots, 8, \infty$ bits from left to right.}\label{figure_3}
\end{figure}

In the first simulation, the cumulative distribution function (CDF) of the achievable rate lower bounds of the users in cell-free and small-cell systems is explored at $C=16, 64, 256$ bps/Hz and $B=1, \dots, 8, \infty$ bits. Figs. \ref{figure_1}, \ref{figure_2}, and \ref{figure_3} correspond to $C=16, 64, 256$ each containing $B=1, \dots, 8, \infty$. According to Fig. \ref{figure_1}, the fronthaul compression act as a bottleneck to the performance of cell-free systems. Regardless of the DAC resolution, cell-free systems are inferior to small-cell systems when $C=16$. In contrast, as the fronthaul capacity increases, Figs. \ref{figure_2} and \ref{figure_3} show that $B\geq 4$ is sufficient for cell-free systems to outperform small-cell systems with infinite-resolution DACs. Therefore, the advantage of cell-free systems can be extracted with low-resolution DACs provided that the fronthaul capacity is sufficient, $C=64$, for example.

\begin{figure}[t]
\centering
\includegraphics[width=1\columnwidth]{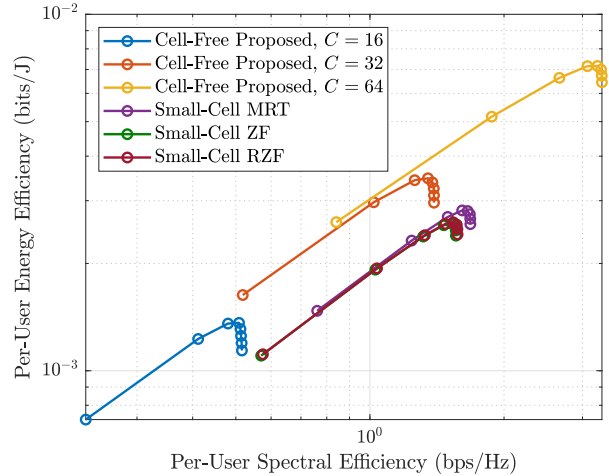}
\caption[caption]{The per-user energy-spectral efficiency lower bound of cell-free and small-cell systems for $C=16, 32, 64$ bps/Hz. At each line, the circles correspond to $B=1, \dots, 8$ bits from left to right.}\label{figure_4}
\end{figure}

In the second simulation, the energy efficiency lower bound of cell-free and small-cell systems is evaluated at $C=16, 32, 64$ bps/Hz and $B=1, \dots, 8$ bits. The energy efficiency lower bound is defined as the spectral efficiency lower bound normalized by the power consumption of the base stations and fronthaul links. The power consumption parameters are adopted from \cite{8333733, 6786060}. According to Fig. \ref{figure_4}, $B=4$ and $B=5$ achieve the maximal energy efficiency lower bound of cell-free systems when $C=32$ and $C=64$, both exceeding that of small-cell systems. Therefore, cell-free systems with low-resolution DACs are energy efficient as long as the fronthaul capacity is not too small, $C=32$, for example.

\section{Conclusion}
In this paper, the ZF precoder with max-min power allocation that accounts for the fronthaul compression and DAC quantization was proposed. The goal of the proposed power allocation was to maximize the minimum achievable rate lower bound of the users given by the AQNM. To solve the max-min power allocation problem, a novel AO method was proposed, whose convergence was shown to be guaranteed. The simulation results showed that moderate fronthaul capacity is sufficient for cell-free systems to outperform small-cell systems.

\section*{Acknowledgment}
This work was partly supported by Institute of Information \& Communications Technology Planning \& Evaluation (IITP) grant funded by the Korea government (MSIT) (No. 2018-0-01410, Development of Radio Transmission Technologies for High Capacity and Low Cost in Ultra Dense Networks) and the National Research Foundation (NRF) grant funded by the MSIT of the Korea government (2019R1C1C1003638).

\bibliographystyle{IEEEtran}
\bibliography{refs_all}

\end{document}